\documentclass{ptapap}
\usepackage{amsmath}
\usepackage{threeparttable}

\author{Melissa Munoz}[QU]
\author{Gregg Wade}[RMC]
\author{Daniel Faes}[SP]
\author{Alex Carciofi}[SP]
\affil[QU]{Department of Physics, Engineering Physics and Astronomy, Queen's University, 64 Bader lane, Kingston, K7L 3N6, ON Canada}
\affil[RMC]{Department of Physics and Space Science, Royal Military College of Canada, PO Box 17000, Stn Forces, Kingston, K7K 7B4, ON, Canada}
\affil[SP]{Instituto de Astronomia, Universidade de São Paulo, Rua do Matão 1226, São Paulo 05508-900, Brazil}

\title{The photometric and polarimetric variability of magnetic O-type stars}

\begin{document}

\maketitle

\begin{abstract}

Massive star winds are important contributors to the energy, momentum and chemical
enrichment of the interstellar medium. Strong, organized and predominantly dipolar magnetic
fields have been firmly detected in a small subset of massive O-type stars. Magnetic massive
stars are known to exhibit phase-locked variability of numerous observable quantities that
is hypothesized to arise due to the presence of an obliquely rotating magnetosphere formed via
the magnetic confinement of their strong outflowing winds.
Analyzing the observed modulations of magnetic O-type stars is thus a key step towards the
better understanding of the physical processes that occur within their magnetospheres. The
dynamical processes that lead to the formation of a magnetosphere are formally solved utilizing
complex MHD simulations. Recently, an Analytic Dynamical Magnetosphere (ADM) model has
been developed that can quickly be employed to compute the time-averaged density, temperature and
velocity gradients within a dynamical magnetosphere.
Here, we exploit the ADM model to compute photometric and polarimetric observables of magnetic Of?p stars, to test geometric models inferred from magnetometry. We showcase important results on the prototypical Of?p-type star HD 191612, that lead to a better characterization of massive star wind and magnetic properties.
\end{abstract}

\section{Introduction}

Magnetic O-type stars are a rare occurrence. Indeed, there are only 11 O-type stars in our Galaxy that are known to host firmly detected magnetic fields \citep{Grunhut2017}. These magnetic fields are  predominantly dipolar and are generally oblique with respect to the rotation axis of the star. As a result, they are known to manifest phase-locked variability of numerous observable quantities, such as their photometric brightness, longitudinal magnetic field strength and line equivalent width. 


In this paper we analyse the photometric and broadband linear polarimetric variability expected of an obliquely rotating magnetic O-type star. By matching models to observations, we can infer or constrain important wind and magnetic properties of a magnetic massive star that are important for the evolution and fate of such stars.  

In section 2, we describe the numerical method that we have designed for photometric and polarimetric synthesis. Next, we attempt to model the observed light curve and Stokes $Q$ and $U$ curves of the prototypical O-type star, HD 191612. We conclude in the final section. 

\section{Numerical method}

\subsection{The ADM model}
The presence of a strong and organized magnetic field can significantly alter the ouftlowing wind structure of a hot star. In fact, under strong magnetic confinement, the closed (presumably dipolar) magnetic field lines channel the winds into the formation of a co-rotating magnetosphere. For slowly rotating O-type stars, the trapped material falls back onto the star on a dynamical timescale, thus forming a so-called dynamical magnetosphere \citep[e.g.][]{Petit2013}. 

\begin{figure}
	\includegraphics[width=0.95\textwidth]{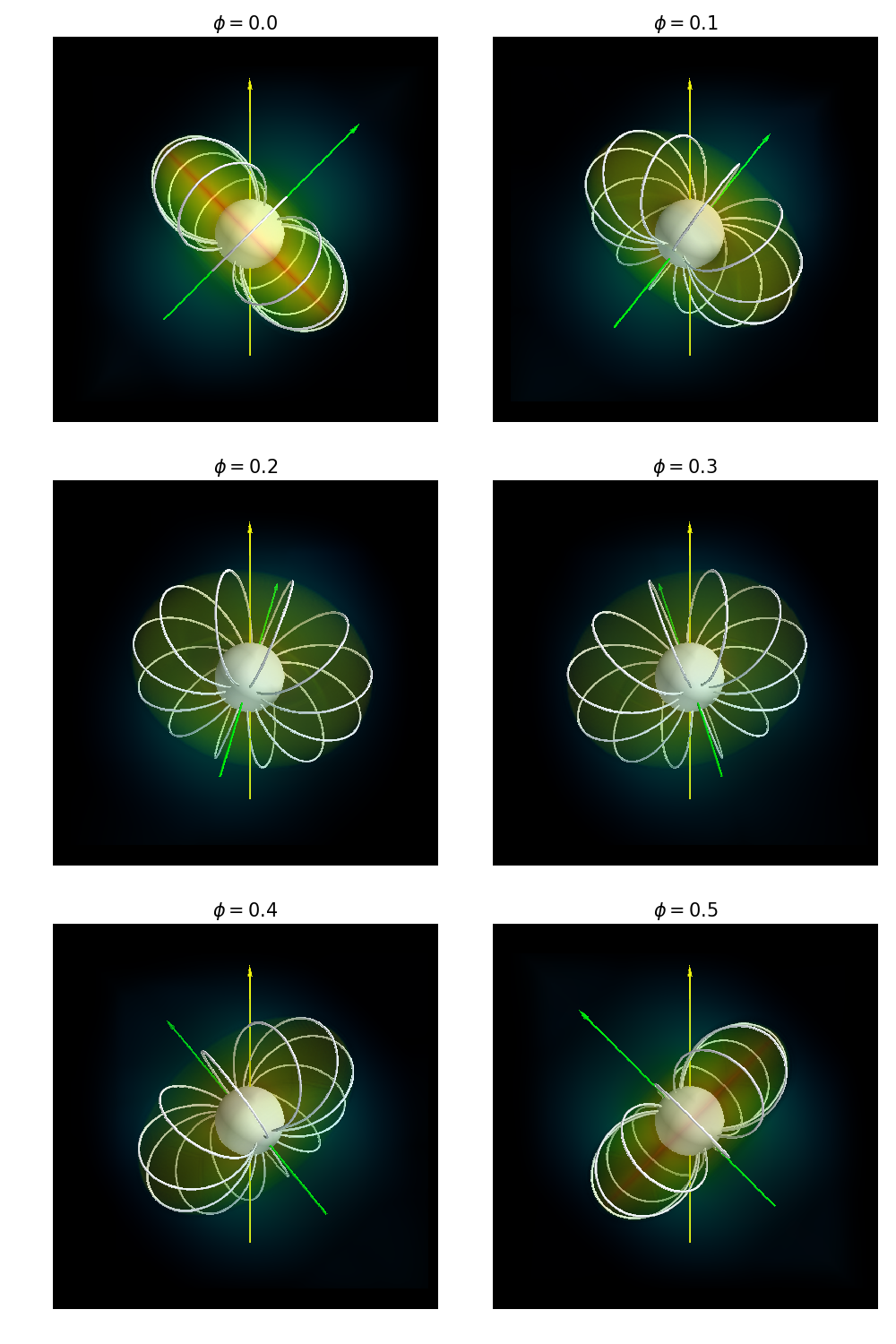}
	\caption{3D rendering of the density structure computed with the ADM model. Red regions are high in density while green regions low in density. The field lines of a dipolar magnetic field are overplotted. The magnetic field axis (inclined green arrow) is tilted by $45^\circ$ with respect to the rotation axis of the star (vertical yellow arrow). One-half of a rotation is illustrated. }
	\label{fig:1}
\end{figure}

The dynamics of a such a magnetosphere can be formally understood using sophisticated 2D or 3D magnetohydrodynamic simulations \citep{UdDoula2002,UdDoula2008,UdDoula2009,UdDoula2013}. However, an analytical model has recently been developed by \citet{Owocki2016} that can quickly compute the steady-state temperature, velocity and density structure (in 2D) of a dynamical magnetosphere. This analytical dynamical magnetosphere (ADM) model essentially serves as a time-averaged picture of the MHD simulations all while being easily adaptable to different stellar and magnetic properties. The input parameters that the ADM model requires are: the mass-feeding rate ($\dot{M}_{B=0}$), polar magnetic field strength ($B_d$), wind terminal velocity ($v_\infty$), stellar effective temperature ($T_\text{eff}$), stellar mass ($M_*$), and stellar radius ($R_*$). 
In the following we exploit the ADM model as a means to simulate the variability resulting from magnetospheres of magnetic O-type stars. To model an oblique magnetic rotator, we rotate the 2D results from ADM into a 3D grid (assuming azimuthal symmetry) and tilt the magnetic axis with respect to the rotation axis. Fig. \ref{fig:1} illustrates snapshots of the magnetosphere density structure computed from ADM viewed at several 
rotational phases ($\phi={0.0, 0.1, 0.2, 0.3, 0.4, 0.5}$). The magnetic axis is tilted by $45^\circ$ with respect to the rotation axis of the star and the density structure is 
computed with input parameters based on the typical prototypical magnetic massive star HD 191612: $T_\text{eff}=35$\,kK, $R_*=14.5$\,$R_*$, $M_*=30$\,$M_*$, $v_\infty = 2700$\, km s$^{-1}$, $\dot{M}_{B=0}=10^{-6}$\,$M_\odot$yr$^{-1}$ and $B_\text{d} = 2.5$\,kG. We can see that  magnetic confinement results in the formation of an overdensity region near the magnetic equator. 

\subsection{The photometric and polarimetric model }

The presence of an obliquely rotating magnetosphere is suspected to be responsible for the modulations of the observable quantities of  magnetic massive stars. As the star rotates, its own magnetosphere periodically occults the source star's light. For hot stars with winds primarily dominated by the electron scattering opacity, the bulk of their photometric and polarimetric variability can be estimated under the single-electron scattering approximation. In this case, the photometric variability is determined by the column density, while the polarimetric variability is characterised by the general shape of the magnetosphere. Both of these quantities can be estimated by utilizing the ADM model to provide the density structure of the magnetosphere.

\begin{figure} 
	\includegraphics[trim={0 1cm 0 0},clip,width=0.85\textwidth]{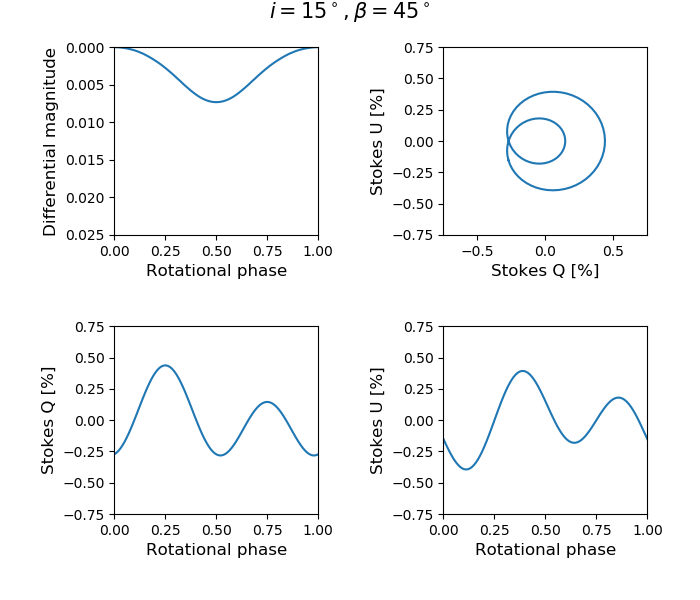}
	\caption{Mosaic of modelled photometric and polarimetric observable quantities for  $i=15^\circ$ and $\beta=45^\circ$. }
	\label{fig:2}
\end{figure}
\begin{figure}
	\includegraphics[trim={0 1cm 0 0},clip,width=0.85\textwidth]{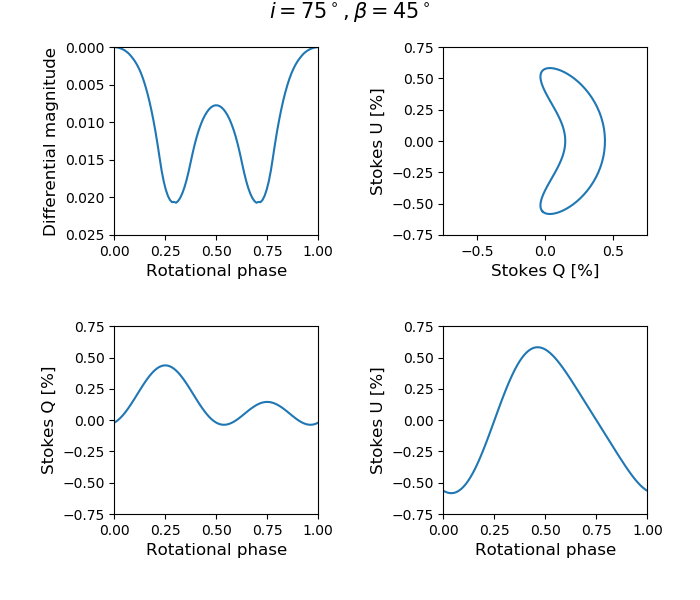}
	\caption{Same as Fig.\ref{fig:2} but for $i=75^\circ$ and $\beta=45^\circ$. }
	\label{fig:3}
\end{figure}

When observations are considered, it is important to consider the inclination angle of the stellar rotation axis ($i$), in addition to the magnetic obliquity ($\beta$). Figs. \ref{fig:2} and \ref{fig:3} display predicted photometric and polarimetric rotational modulations under two configurations: $i=15^\circ$, $\beta=45^\circ$ and $i=75^\circ$, $\beta=45^\circ$. The stellar and magnetic parameters are set to mimic those of HD 191612:  $T_\text{eff}=35$\,kK, $R_*=14.5$\,$R_\odot$, $M_*=30$\,$M_\odot$ and $v_\infty = 2700$\, km s$^{-1}$ and $B_\text{d} = 2.5\pm0.4$\,kG. We can see that the photometric light curves are either single dipped (if $i+\beta < 90^\circ$) or double dipped (if $i+\beta > 90^\circ$).  In contrast, the broadband Stokes $Q$ and $U$ curves are generally sinusoidal. In $Q-U$ space, the linear polarimetric variability appears as a single looped locus (if one of Stokes $Q$ or $U$ curve is single waved) or double looped (if both Stokes $Q$ and $U$ curves are double waved). 

\section{Application}

HD 191612 is a well-studied, slowly rotating Of?p-type star. With a rotational period of $\sim537$\,d and a magnetic field strength of $\sim2.5$\,kG, periodic modulations can be observed in its photometric brightness, longitudinal magnetic field strength and $H\alpha$ equivalent width \citep{Wade2011}. In the following, we attempt to model both the photometric and polarimetric variability of this star.

\subsection{Fit to the photometry of HD 191612}

\begin{figure}
	\includegraphics[width=1.0\textwidth]{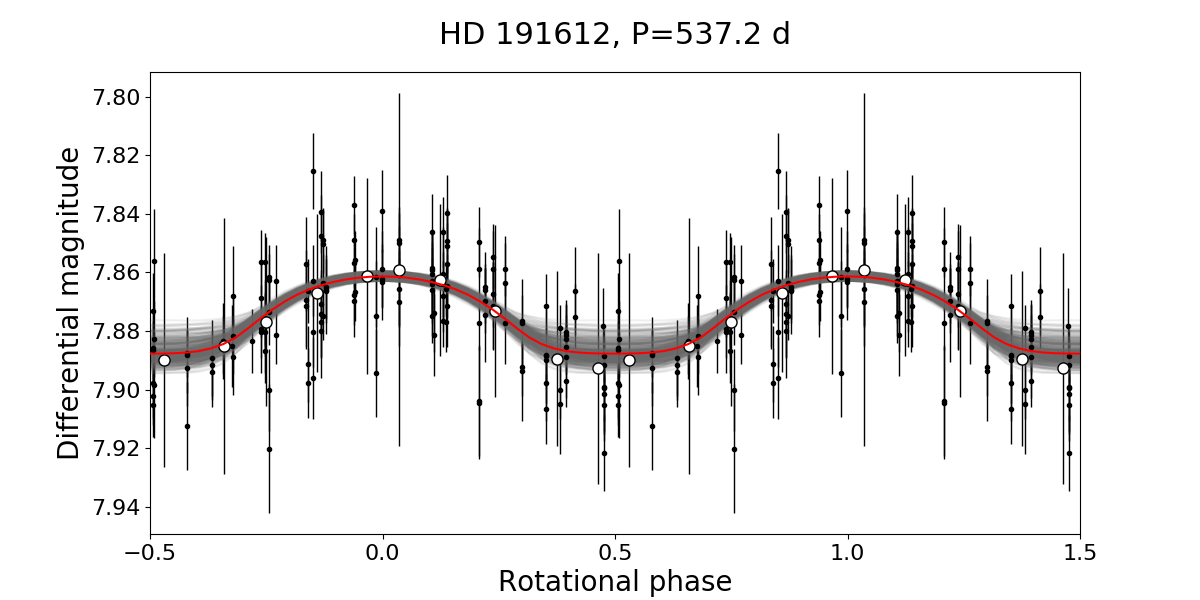}
	\caption{The phased Hipparcos photometric variability of HD 191612. The curve of best-fit is overplotted in red (bold solid lines). Curves that span the 1 $\sigma$ error bars on the best-fit parameters are overplotted in grey (thin solid lines).}
	\label{fig:4}
\end{figure}

The photometry for HD 191612 was retrieved from the Hipparcos archive, and were originally obtained between November 1989 and February 1993. The light curve was phased according to the period and ephemeris derived by \citet{Wade2011} utilizing a combination of all archival $H\alpha$ equivalent width measurements (see Fig. \ref{fig:4}): JD $= 2,453,415.1(5) \pm 537.2(3) \times E$.  

To model the Hipparcos light curve for HD 191612, we fix its stellar parameters to those obtained by \citet{Howarth2007}. In addition, to further constrain the mass-feeding rate, we set the polar field strength to its inferred value of $2.5\,kG$ (from  longitudinal magnetic field measurements \citep[see][]{Wade2011}) . The curve of best fit is shown in Fig. \ref{fig:4} with best-fit parameters listed in Table \ref{tab:1}. We obtain a mass-feeding rate of $\log \dot{M}_{B=0}=-5.72_{-0.05}^{+0.13}$ and two possible magnetic geometries: ${i=27^{+13}_{-14}}^\circ$ and ${\beta=61^{+13}_{-11}}^\circ$, or $
{\beta=27^{+13}_{-14}}^\circ$ and ${i=61^{+13}_{-11}}^\circ$.


\subsection{Fit to the polarimetry of HD 191612}

Ground-based linear polarimetry of HD 191612  was obtained from the IAGPOL polarimeter  mounted on the Boller \& Chivens telescope at OPD/LNA, Brazil. The data were taken between June 2011 and August 2016.

\begin{figure}
	\includegraphics[width=1.0\textwidth]{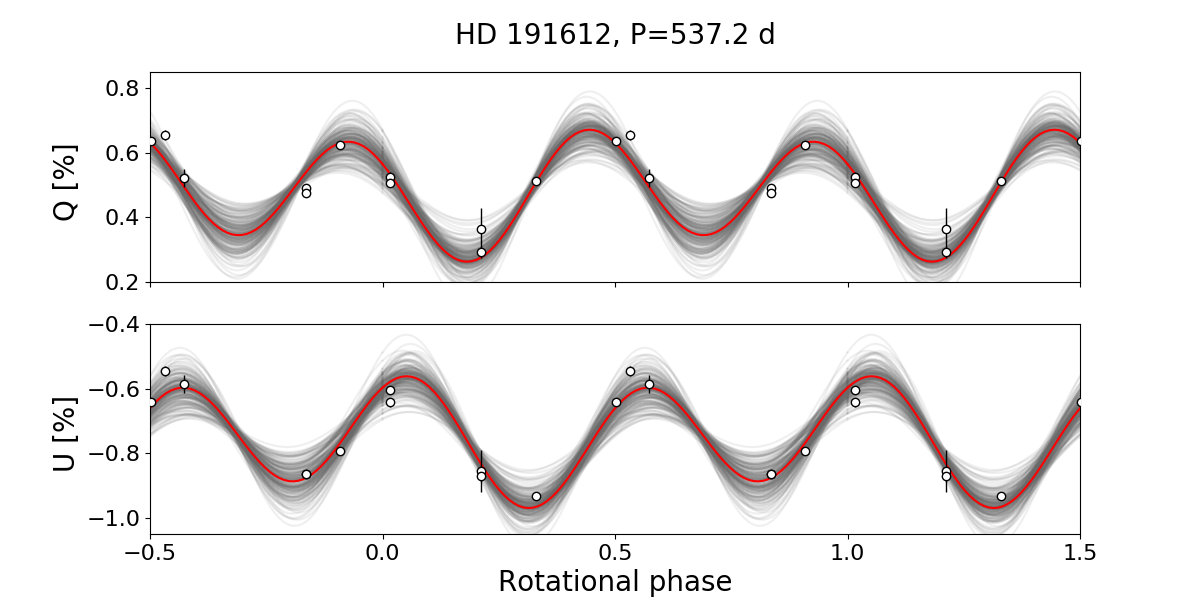}
	\caption{The phased polarimetric variability of HD 191612 (From top to bottom: Stokes $Q$ and $U$ curves). The curve of best-fit is overplotted in red (bold solid lines). Curves that span the 1 $\sigma$ error bars on the best-fit parameters are overplotted in gray (thin solid lines).}
	\label{fig:5}
\end{figure}

The polarimetric fitting is accomplished similarly to the photometric modelling. We fix the stellar properties to the known values for this star and in addition fix its detected magnetic polar field strength. We model the Stokes $Q$ and $U$ curves simultaneously (see Fig. \ref{fig:5}) and constraining the magnetic geometry to ${i=21^{+22}_{-12}}^\circ$, ${\beta=68^{+10}_{-19}}^\circ$ with a mass-feeding rate of $\log \dot{M}_{B=0}=-5.79^{+0.13}_{-0.07}$ (see Table \ref{tab:2}).

\section{Discussion}

\begin{table}
	\begin{center}
		\caption{Best-fit parameters to the Hipparcos photometry of HD 191612}
		\label{tab:1}
		\begin{tabular}{ccccccc} 
			\hline
			Star &$i+\beta$&$|i-\beta|$&  $i$ or $\beta$ & $i$ or $\beta$ & $\log \dot{M}_{B=0}$ & c\\
			&[deg]&  [deg]  &[deg]&[deg]&[$\dot{M}_\odot$ yr$^{-1}$]  &[mmag]\\
			\hline
			HD 191612&$88_{-5}^{+5}$& $33_{-23}^{+26}$& $27_{-14}^{+13}$& $61_{-11}^{+13}$ & $-5.72_{-0.05}^{+0.13}$ & $7861_{-2}^{+2}$ \\
			\hline
		\end{tabular}
	\end{center}
	\begin{tablenotes}
		\small
		\item $^\dagger$ $\Delta m_0$ corresponds to a vertical offset in the differential magnitude (assumed constant).
	\end{tablenotes}	
\end{table}

\begin{table}
	\begin{center}
		\caption{Best-fit parameters to the polarimetry for HD 191612}
		\label{tab:2}
		\begin{tabular}{cccccc} 
			\hline
			Star & $i$ & $\beta$ & $\log \dot{M}_{B=0}$ & $Q\textsubscript{IS}^\dagger$ &  $U\textsubscript{IS}^\dagger$  \\ 
			& \text{[deg]}   & [deg] &[$\dot{M}_\odot$ yr$^{-1}$]         &           [\%]        &[\%]                    \\
			\hline
			HD 191612 & $21_{-12}^{+22}$& $68_{-19}^{+10}$ & $-5.79_{-0.07}^{+0.13}$ & $-0.746_{-0.012}^{+0.012}$ & $-0.046_{+0.013}^{-0.009}$\\
			\hline
		\end{tabular}
	\end{center}
	\begin{tablenotes}
		\small
		\item $^\dagger$ $Q\textsubscript{IS}$ and $U\textsubscript{IS}$ respectively correspond to the Stokes $Q$ and $U$ interstellar linear polarisation (assumed constant).
	\end{tablenotes}
\end{table}

In the previous section, we were successful in reproducing both the photometric and polarimetric variability of HD 191612. Our derived magnetic geometry is consistent with the tentative results from \citet{Wade2011}. Implied from the longitudinal magnetic field  modulations, their magnetic geometry was constrained to the family of solutions satisfying ${i+\beta=95^\circ\pm10}^\circ$. Similarly, we obtained  ${i+\beta=88^{+5}_{-5}}^\circ$ from the photometry and ${i+\beta=89^{+16}_{-15}}^\circ$ from the linear polarisation. 

The ambiguity present in the geometric angles determined via the magnetic variations also affects the inferences from the photometric variations. This is an expected result as the modelling of these two observable quantities are degenerate to interchanges in the $i$ and $\beta$ angles. As a result, they cannot be distinguished. However, the quantity that can be more reliably constrained is in fact the $i+\beta$ sum. In contrast, the magnetic geometry derived from the polarimetric variations does not suffer from the same degeneracy. In fact, the $i$ and $\beta$ angles can instead be constrained independently. For HD 191612, we obtained ${i=21^{+22}_{-12}}^\circ$ and ${\beta=68^{+10}_{-19}}^\circ$. While still satisfying the $i+\beta=95^\circ+10^\circ$ condition \citep[derived from][]{Wade2011}, the geometry inferred from the polarimetry clearly favours one of the two possible geometries inferred from the photometry. This is a natural benefit of linear polarisation modelling.


Moreover, the mass-feeding rate obtained via photometric and polarimetric modelling are also in agreement. We stress that the mass-feeding rate encoded in the ADM model is in fact a hypothetical mass-loss rate in absence of a magnetic field. Wind quenching due to the presence of a magnetic field can  effectively decrease the true measured mass-loss rate. According to \citet{UdDoula2002}, we can estimate the mass-loss rate via the scaling relation $\dot{M}=f_\text{B} \dot{M}_\text{B=0}$, where $f_{B}$ is the magnetic factor. For HD 191612, we obtain a magnetic factor of $\sim0.19$ and therefore a reduced mass-loss rate of $\dot{M}=10^{-6.44}$\,$\dot{M}_\odot$ yr$^{-1}$ and $\dot{M}=10^{-6.51}$\,$\dot{M}_\odot$ yr$^{-1}$ inferred from the photometric and polarimetric modelling respectively. Via $H\alpha$ diagnostics, a clumped mass-loss rate of $10^{-5.8}$\,$\dot{M}_\odot$ yr$^{-1}$ was derived by \citet{Howarth2007}. For this value to be compatible with our results, a clumping factor in the order of $\sim25$ is necessary. This is in agreement with typical clumping factors for these stars that are suspected to range between 2 and 100 \citep{Puls2008}.

\section{Conclusion}
We have described a numerical model capable of synthesising the photometric and polarimetric variability of magnetic hot stars due to magnetospheric scattering. We have applied our model to reproduce the rotational modulations observed in the photometry and polarimetry of HD 191612. Our modelling results constrain the magnetic geometry and mass-feeding rate to ${i=27_{-14}^{+13}}^\circ$, ${\beta=61_{-11}^{+13}}^\circ$ and $\log \dot{M}_{B=0}=-5.72_{-0.05}^{+0.13}$ from the photometry and to ${i=21_{-12}^{+22}}^\circ$, ${68_{-19}^{+10}}^\circ$ and $\dot{M}_{B=0}=-5.79_{-0.07}^{+0.13}$ from the linear polarisation. A natural next step is to conduct a simultaneous fit of numerous observable quantities as a means to further constrain these parameters. 

\bibliographystyle{ptapap}
\bibliography{munoz}

\end{document}